\documentclass[3p,times,twocolumn]{revtex4}
\usepackage{amssymb}
\usepackage[figuresright]{rotating}

\newcommand{\kt}{\ensuremath{k_{\mathrm{T}}}}
\newcommand{\pt}{\ensuremath{p_{\mathrm{T}}}}
\newcommand{\ptd}{\ensuremath{p_{\mathrm{T,D}}}}
\newcommand{\ptjet}{\ensuremath{p_{\mathrm{T,jet}}}}
\newcommand{\ptjetch}{\ensuremath{p_{\mathrm{T,jet}}^\mathrm{ch}}}
\newcommand{\GeVc}{\ensuremath{\mathrm{GeV}/c}}
\newcommand{\mjet}{\ensuremath{M_{\mathrm{jet}}}}
\newcommand{\raa}{\ensuremath{R_\mathrm{AA}}}
\begin{document}

%\begin{frontmatter}




\title{Jet Fragmentation and Jet Shapes in JEWEL and q-PYTHIA}


\author{M. van Leeuwen}
\affiliation{Nikhef, National Institute for Subatomic Physics, P.O. Box 41882, 1009 DB  Amsterdam, and Utrecht University, P.O. Box 80000, 3508 TA Utrecht, The Netherlands}

%\begin{keyword}
%jet fragmentation \sep heavy-ion collisions
%\PACS 25.75.-q \sep 25.75.Bh \sep 13.87.Fh 

%\end{keyword}

%\end{frontmatter}

% \linenumbers

\begin{abstract}
A study of several observables characterising fragment distributions of medium-modified parton showers using the JEWEL and Q-PYTHIA models is presented, with emphasis on the relation between the different observables.
\end{abstract}

\maketitle

%\section{Models and analysis setup}
We explore the mechanism for parton energy loss in a Quark Gluon Plasma by comparing and contrasting two event generators with medium-modified parton showers: JEWEL \cite{Zapp:2013vla} and Q-PYTHIA \cite{Armesto:2009fj} and confronting them with experimental data. Measurements at LHC have shown that jet rates are suppressed with respect to expectations from an independent superposition of nucleon-nucleon collisions (nuclear modification factor $\raa < 1$) indicating significant out-of-cone radiation, while only relatively small modifications of the fragment distributions in the jet cone are found. We explore whether the models can reproduce these effects.

The JEWEL event generator simulates in-medium shower evolution based on an initial state from PYTHIA \cite{Sjostrand:2006za} and elastic interactions between the hard initial partons in the shower and medium partons. Repeated interactions induce gluon radiation. Interference effects between subsequent emissions are modeled using formation times, which has been shown to agree with analytical calculations in the appropriate limits.% \cite{Zapp:2012ak}. 
The JEWEL code is publicly available and the relation between medium density and collision energy has been tuned to reproduce charged particle \raa{} measurements at RHIC and LHC. The medium density profile is based on a longitudinally expanding Glauber overlap; the local temperature is sampled to determine the density of scattering centers and their momentum distribution.

The Q-PYTHIA event generator is also based on PYTHIA and models medium-induced radiation by an enhanced splitting probability.
%, based on the quenching weights from Salgado and Wiedemann \cite{Salgado:2003gb}. 
The version of Q-PYTHIA used here is interfaced to an optical Glauber model which is sampled 
%using simulated parton trajectories along which the medium density is integrated 
to calculate an equivalent density for a uniform medium for each parton passing through the medium, as used in the PQM model \cite{Dainese:2004te}
%,Salgado:2002cd}. 
This version of Q-PYTHIA is available as part of the AliRoot software package \cite{aliroot}. The overall scale parameter for the medium density profile is set to $k=1.8\cdot10^{6}$ which gives $\raa \approx 0.2$, similar to the measured charged particle \raa{} at LHC.

The final fragmentation and hadronisation of the medium modified shower is handled by PYTHIA in both JEWEL and Q-PYTHIA.

The analysis uses all final state charged and neutral particles, except neutrinos, which are not detected in the experiments. Jets are reconstructed with the anti-\kt{} algorithm from Fastjet \cite{Cacciari:2011ma} with the \pt-scheme. The results are given for pseudo-rapidity range $|\eta| < 1$ for the spectra and $|\eta| < 2$ for the fragment distributions. For the fragment distributions, only charged particles are used, while the jet energy is based on charged and neutral fragments, like in the experiments.

\begin{figure*}
\begin{center}
\includegraphics[width=0.45\textwidth]{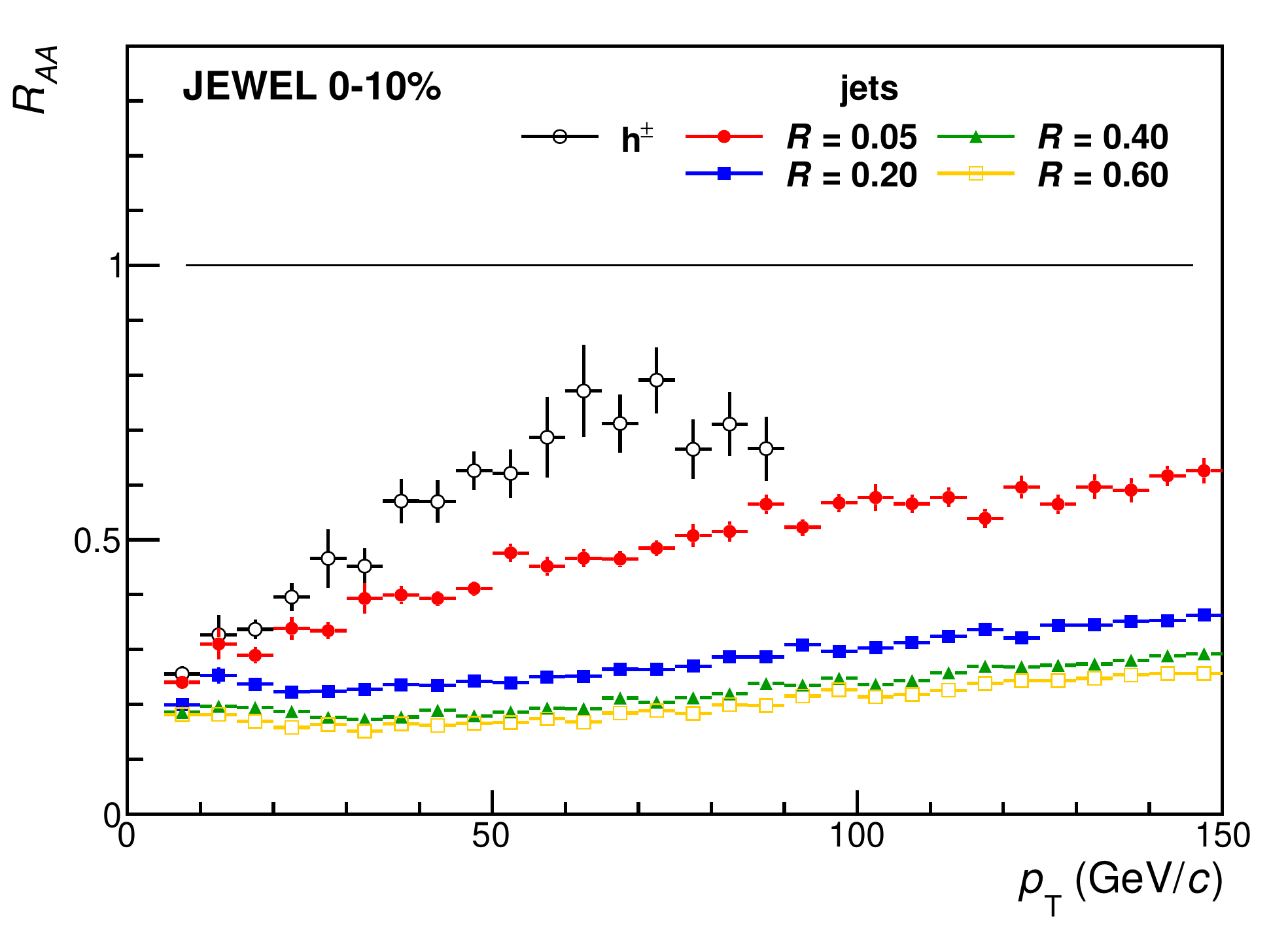}
\includegraphics[width=0.45\textwidth]{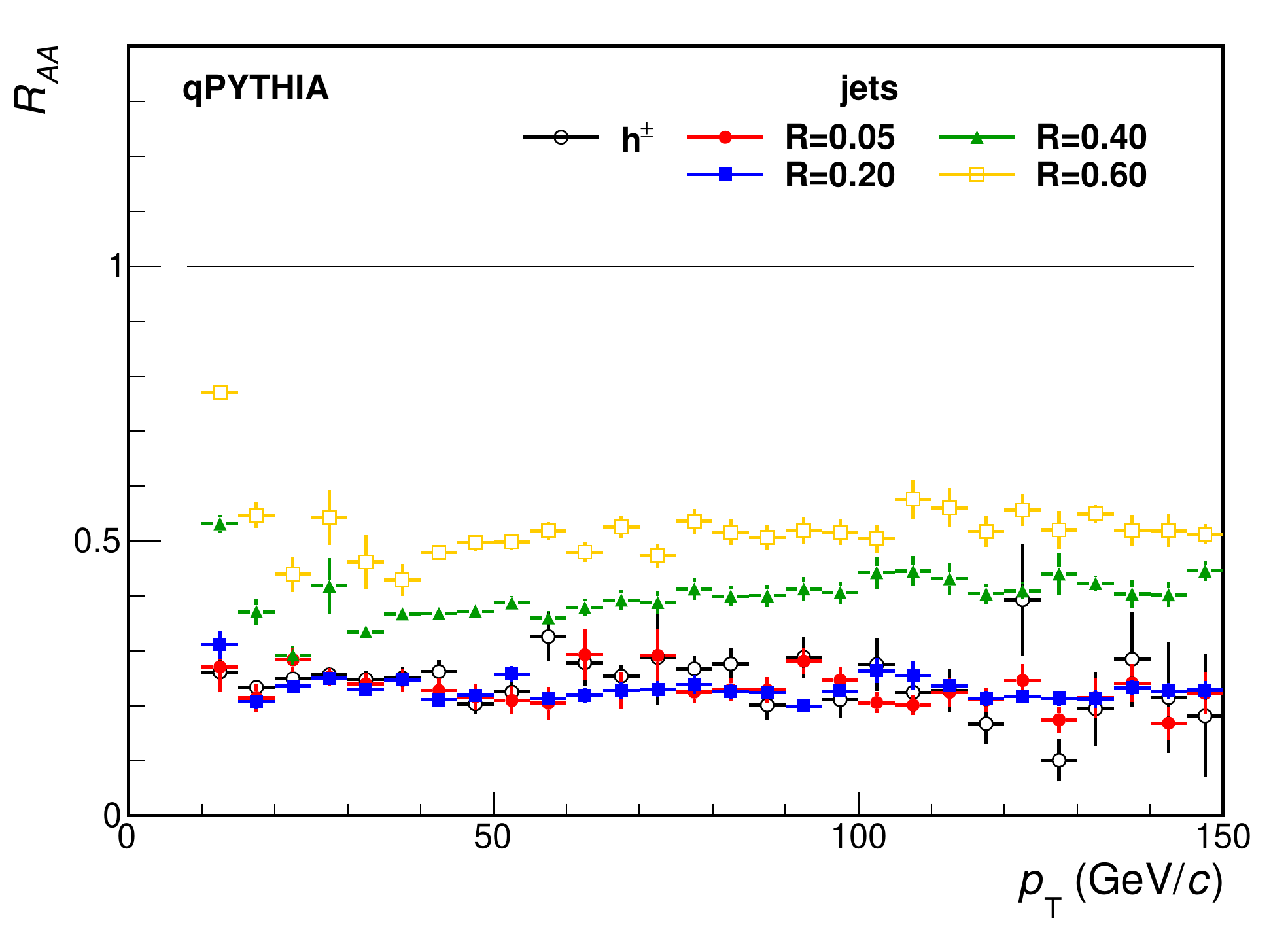}
\end{center}
\caption{\label{fig:raa_r}Nuclear modification factor \raa{} for charged particles and anti-\kt{} jets with different resolution parameter $R$ from JEWEL (without recoil hadrons) and Q-PYTHIA for Pb--Pb collisions with $\sqrt{s_{NN}}=2.76$~TeV.} 
\end{figure*}

\section{Recoil in JEWEL and background subtraction}
In JEWEL, momentum exchange between the jet and the medium takes place via 2-to-2 scatterings with medium partons. By default, JEWEL only keeps track of the shower partons, while the recoiling medium parton is not kept in the event record. JEWEL can also be used a mode where the recoil partons are kept in the event. In this mode, a much larger number of particles is generated. Some of these particles are hadronisation products of medium partons, with an approximately uniform azimuthal distribution. The median-based background subtraction from Fastjet was used to remove the contribution of medium fragments from the jet energy, using a strip of width $2R$ in pseudo-rapidity around the jet under study to measure the background with the \kt{} jet algorithm. In addition, for the longitudinal and transverse fragment distributions, the contribution to the fragment distribution was estimated by measuring particles in cones at 90 degrees with respect to the jet under study and this contribution is subtracted. For the jet shape variables (\ptd{} and radial moment $g$), two different background subtraction methods were used: the derivative method \cite{Soyez:2012hv} and the constituent subtraction method \cite{Berta:2014eza}.

\section{Results and comparison to existing data}
\label{}

Figure \ref{fig:raa_r} shows the nuclear modification factor calculated in Q-PYTHIA and JEWEL, for charged hadrons and for jets with different resolution parameter $R$. The dependence of \raa{} on the resolution parameter is seen to be quite different for the two models: Q-PYTHIA shows not much change for $R<0.2$ and then a monotonic increase with $R$, while JEWEL shows a rapid decrease with increasing $R$ for small values of $R<0.2$ and then a slower decrease. In addition, JEWEL shows an increase of \raa{} with \pt{} for charged particles which is compatible with what is observed in the experiments, while \raa{} in Q-PYTHIA is rather independent of \pt. 

\begin{figure}
\begin{center}
\includegraphics[width=0.4\textwidth]{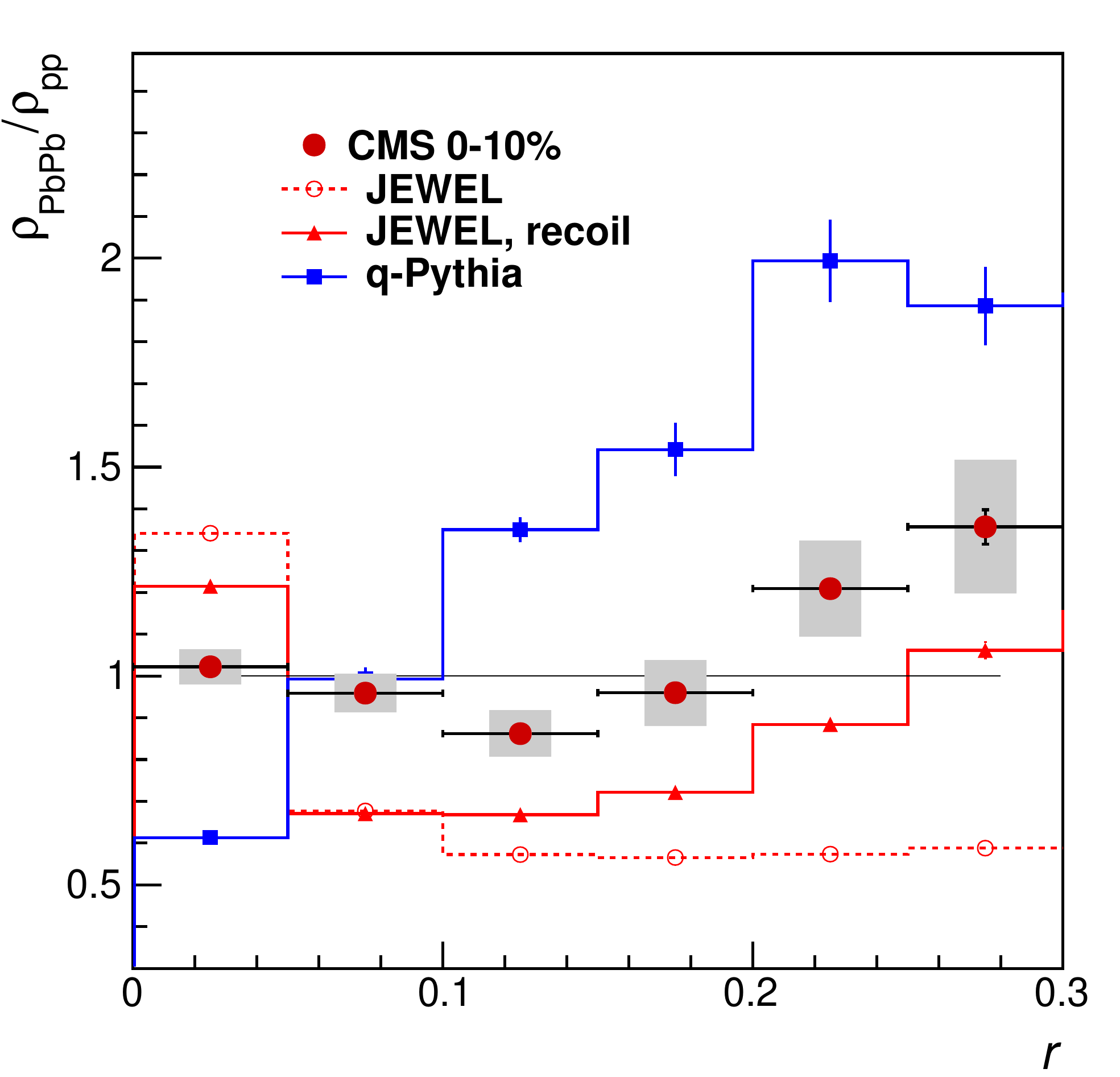}
\end{center}
\caption{\label{fig:rho}Ratio of jet profiles in Pb--Pb and pp collisions in jets with $R=0.3$ from JEWEL (with and without recoil hadrons) and Q-PYTHIA. The red circle markers indicate the results from CMS. \cite{Chatrchyan:2013kwa}} 
\end{figure}

Figure \ref{fig:rho} shows the modifications of the jet momentum flow profile
\[
\rho(r \pm \delta r) = C \frac{1}{2 \delta r}\sum_{i, r_i \in [r-\delta r, r+\delta r]} {\frac{p_{\mathrm T, i}}{\ptjet}}
\]
which is the sum of the transverse momentum of the jet constituents $p_{\mathrm{T},i}$ as a function of the distance to the jet axis $r=\sqrt{\Delta \eta^2 + \Delta \phi^2}$. The profile is averaged over all jets in the \pt-range under study: $100 < \ptjet < 120$ \GeVc. The normalisation constant $C$ normalises the integral over the profile to unity, which corrects for the fact that the profile is measured using charged tracks, while the jet energy is based on charged and neutral constituents. Figure \ref{fig:rho} shows the ratio of the profiles in Pb-Pb and pp collision measured by CMS \cite{Chatrchyan:2013kwa}, compared to results from JEWEL and Q-PYTHIA. It can be seen in the figure that JEWEL has a significant increase of the momentum flow close to the jet axis, followed by a depletion for $r>0.05$. At larger $r$, recoil hadrons increase the momentum flow, while JEWEL without recoil shows a constant depletion, up to $r=0.3$. There is a clear relation between the radial profile measurement and the dependence of jet \raa{} on the resolution parameter $R$ as shown in Fig.\ \ref{fig:raa_r}: the strong decrease of $\raa$ with $R$ at small radii is related to the slope of $\rho(r)$ vs $r$; at larger $R$, \raa{} still decreases because $\rho(r)$ is smaller in Pb--Pb than in pp collisions.
For Q-PYTHIA, a rising trend is seen in $\rho(r)$ which is related to the increase of \raa{} with $R$. The trend in the JEWEL calculation is closer to the measured behaviour.

It should be noted that in the CMS measurement, no correction is applied for the effect of background fluctuations in the Pb--Pb result, but fluctuations have rather been applied to the pp result to obtain a valid comparison between pp and Pb-Pb. In the JEWEL calculations, the fluctuations were not applied, but it is expected that the effect cancels out to first order in ratios.

\begin{figure}
\begin{center}
\includegraphics[width=0.4\textwidth]{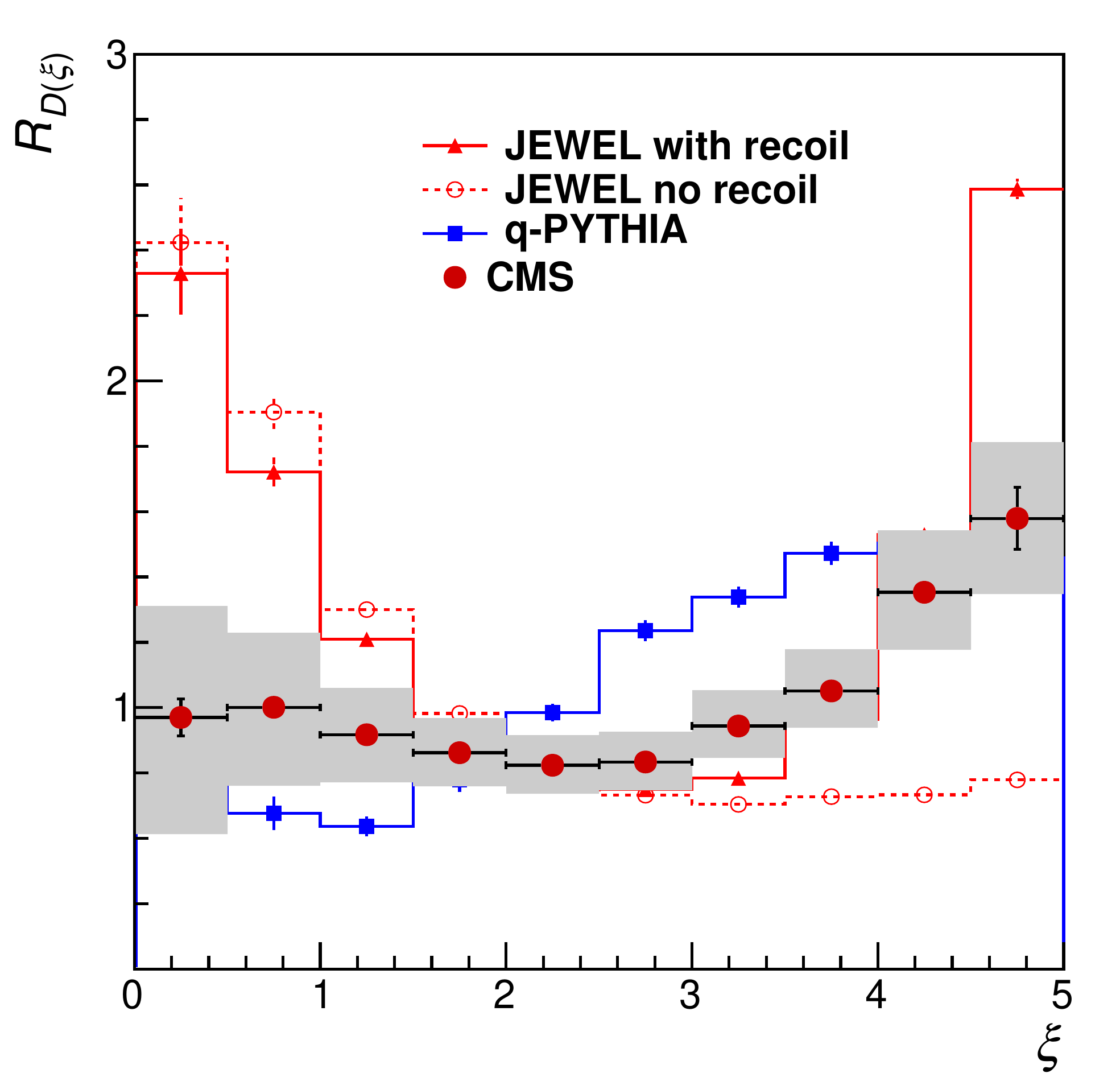}
\end{center}
\caption{\label{fig:xifrag}Ratio of the longitudinal fragment distribution in Pb--Pb and pp collisions as a function of $\xi=-\log(\pt/\ptjet)$ from JEWEL, Q-PYTHIA, and measurements from CMS (red circle markers). \cite{Chatrchyan:2014ava}} 
\end{figure}

Figure \ref{fig:xifrag} shows the modification of the longitudinal fragment distribution $D(\xi)$ as a function of $\xi=-\log{\pt/\ptjet}$. The Q-PYTHIA calculation shows a softening of the fragment distribution: suppression at low $\xi$ (large \pt{}) and an increase at large $\xi$ (low \pt). In JEWEL, a depletion is seen at intermediate $\xi$ and an enhancement for small and large $\xi$. The enhancement at small $\xi$ is related to the increase at small $r$ in the radial profile, while the increase at large $\xi$ is only seen when recoil hadrons are included, and related to the increase at larger $r$ in $\rho$.
The red markers in the figure show the CMS results \cite{Chatrchyan:2014ava} for this observable (ATLAS performed a similar measurement \cite{Aad:2014wha} using peripheral Pb--Pb collisions as a reference).
 Qualitatively, the JEWEL calculation is closer to the measurement, although JEWEL seems to overestimate the increase of the yield at small $\xi$. A similar effect was seen at small $r$ in Fig.\ \ref{fig:rho}. The note about the treatment of background fluctuations as discussed for the radial profile also applies here.

It is interesting to see that JEWEL and Q-PYTHIA show quite different trends: Q-PYTHIA shows a softening and broadening of the fragmentation for all $r$, while JEWEL shows a hardening and narrowing at small $r$, followed by a broadening and softening at larger $r \gtrsim 0.1$. The broadening in JEWEL is only generated when recoil hadrons are generated. In measurements, so far, a clear broadening and softening is seen for moderate to large $r$, while at small $r$ and larger \pt{} very little modification or a very small enhancement is seen. Qualitatively, both JEWEL and Q-PYTHIA show larger changes of the fragment distributions than are seen in the measurements. This may indicate that in medium-modified shower simulations, a significant suppression of jet rates is accompanied by significant redistribution of the fragments inside the jet cone. It is a key phenomenological question whether this relation between in- and out-of-cone radiation is general for this class of models, and if so, what mechanism would be able to generate the observed suppression of jet rates with only small changes to the in-cone distributions. It would also be interesting to measure the radial profiles in narrower bins in $r$, in particular for small $r$, where JEWEL shows a large enhancement.

\section{Jet shape variables}

\begin{figure*}
\includegraphics[width=0.32\textwidth]{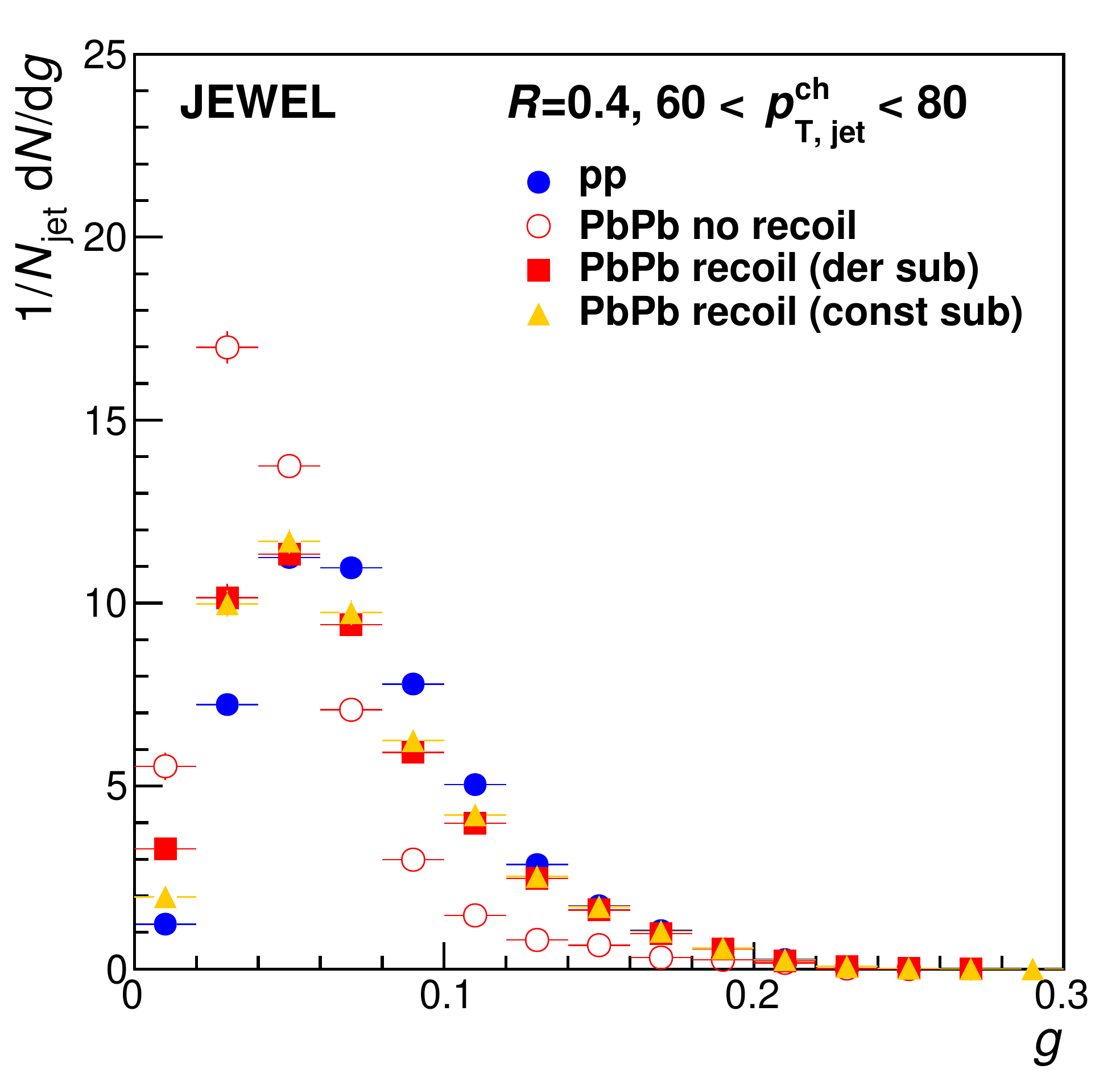}
\includegraphics[width=0.32\textwidth]{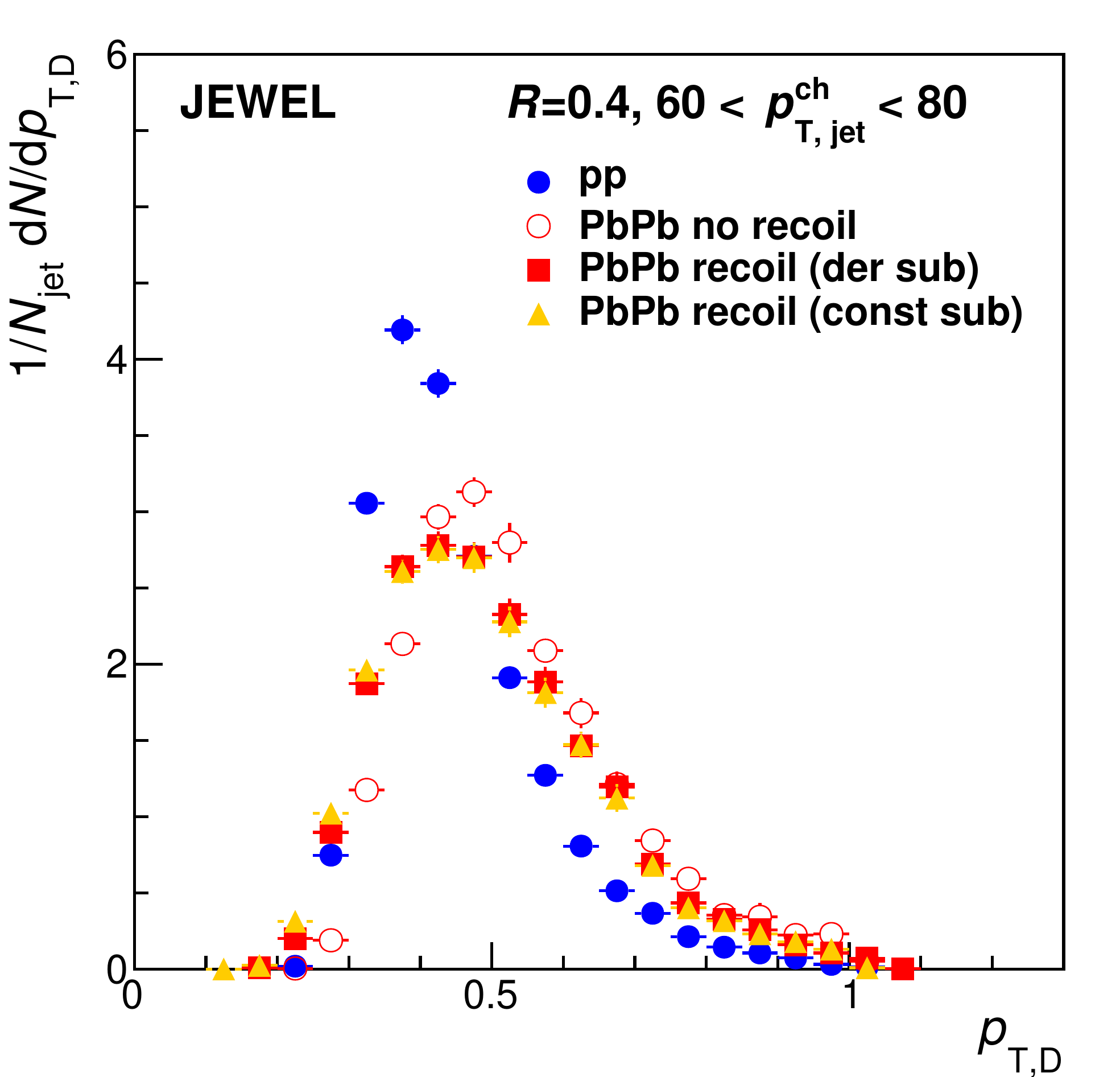}
\includegraphics[width=0.32\textwidth]{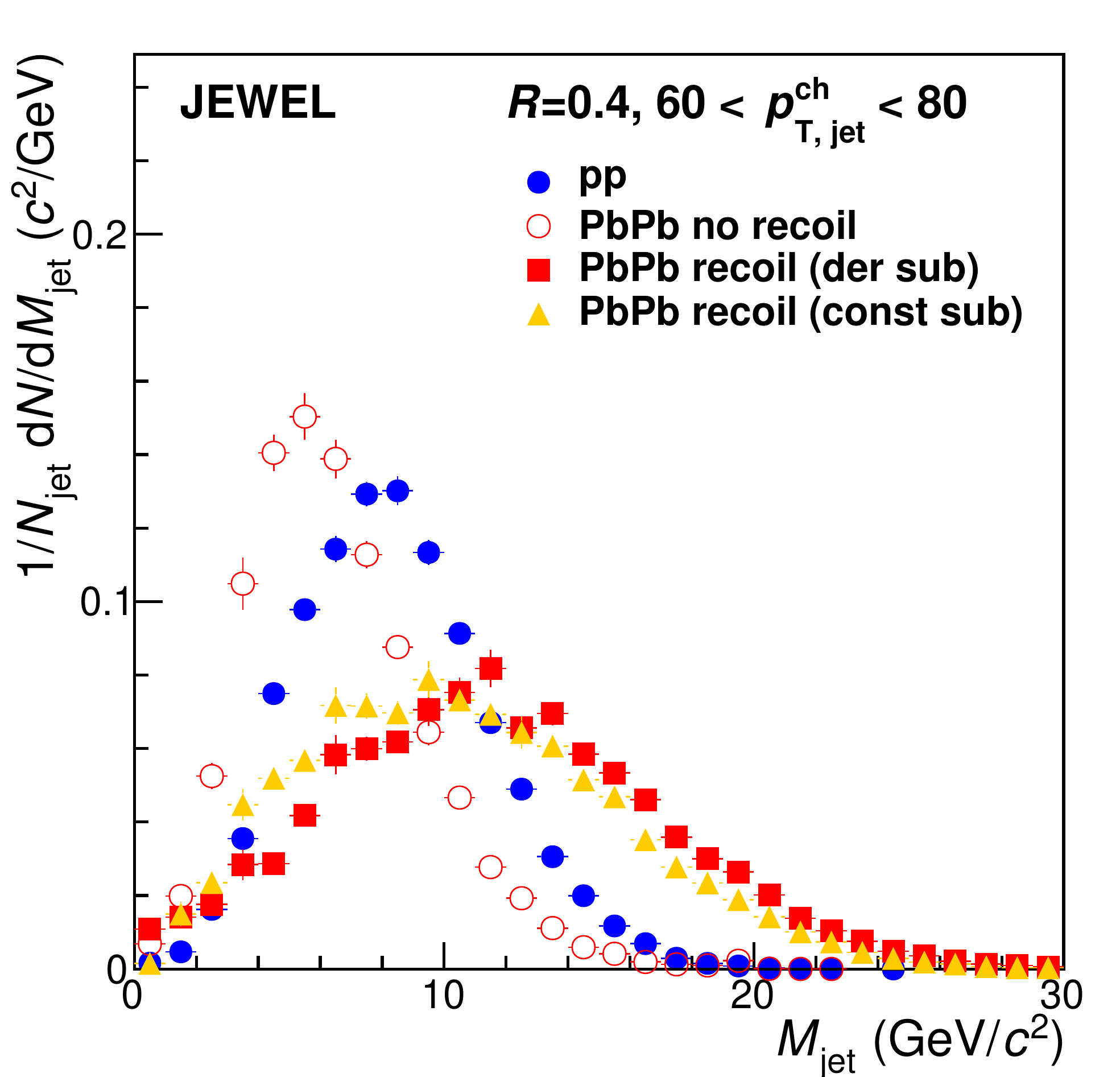}
\caption{\label{fig:shapes}Jet shape distributions for charged jets from JEWEL for pp and Pb--Pb collisions, without and with recoil.} 
\end{figure*}

Jet shape variables provide an alternative measure of fragment distributions in jets. While the  fragment distributions in Figs \ref{fig:rho} and \ref{fig:xifrag} reflect average distributions over many jets, the jet shape variables are calculated jet by jet and the distributions may provide insight in fluctuations of energy loss. Moreover, the correction procedure for underlying event and fluctuations of the underlying event are different, and thus may provide a somewhat independent way to investigate jet structure for in-medium showers. The variety of jet shapes in the literature \cite{Larkoski:2014pca} provides an arsenal of ways to characterise jets, which can be explored for their sensitivity to medium-induced modifications of the jets. A full study of these variables is beyond the scope of these proceedings, so here we 
%For a more complete overview of jet shape variables in the context of quark/gluon separation, we refer the reader to \cite{}. 
explore three jet shapes: the radial moment or girth $g$, the \pt-dispersion \ptd{} which are related to the radial and longitudinal fragment distributions, and the jet mass \mjet{} which is of interest due to its relation to virtuality \cite{Majumder:2014gda}. The girth and \ptd{} are defined as:
\[
g = \sum_i \frac{p_{\mathrm{T},i}}{\ptjet} r_i  \quad \mathrm{and} \quad \ptd = \frac{\sqrt{\sum_i {p_{\mathrm{T},i}^2}}}{\ptjet}.
\]

Figure \ref{fig:shapes} shows these jet shape variables for JEWEL (Q-PYTHIA omitted due to space constraints) for charged particle jets with $60 < \ptjetch < 80$ \GeVc{} (using $E$-scheme recombination). The radial moment $g$ is seen to decrease due to medium modifications; this effect is stronger without recoil than with recoil. This is consistent with the observed narrowing of the jet core in Fig.\ \ref{fig:rho}. For \ptd, the opposite is seen: the distribution shifts to larger values due to medium effects, which is qualitatively consistent with the observed increase of the fragment yield at large $\xi$.

The jet mass \mjet{} is sensitive to both longitudinal and transverse dynamics in the jet (see \cite{Larkoski:2014pca} for a more quantitative discussion). In JEWEL, the jet mass decreases due to medium effects if no recoil hadrons are taken into account, while it increases when recoil hadrons are generated. The present investigation of jet shapes is only meant as a first illustration of what one might expect to see. It will be interesting to further explore this, both in measurements and in models, for example to investigate the sensitivity of such observables to fluctuations in the energy loss and other aspects of energy loss models. 

\section*{Acknowledgments}
The author would like to thank Marta Verweij and Leticia Cunqueiro for bringing jet shape variables to his attention, and Marta Verweij for useful discussions of the initial results and advice on the background subtraction techniques for jet shapes.


%% References

\nocite{*}
\bibliographystyle{elsarticle-num}
\bibliography{VanLeeuwen_M}






% \end{thebibliography}

\end{document}